\documentclass[aps,prl,reprint,superscriptaddress,floatfix,nofootinbib]{revtex4-2}

\usepackage{graphicx}
\usepackage{amsmath}
\usepackage[caption=false]{subfig}
\usepackage{siunitx}
\usepackage{placeins}
\usepackage{color}
\usepackage{standalone}
\usepackage{dcolumn}
\usepackage{tensor}
\usepackage{bm}
\usepackage{microtype}
\usepackage{etoolbox}
\usepackage{amssymb}
\usepackage{mathrsfs}
\usepackage{accents}
\usepackage[normalem]{ulem}
\usepackage[dvipsnames]{xcolor}
\usepackage[colorlinks,urlcolor=NavyBlue,citecolor=NavyBlue,linkcolor=NavyBlue,pdfusetitle]{hyperref}
\usepackage[all]{hypcap}
\usepackage[inline]{enumitem}
\usepackage[utf8]{inputenc}
\usepackage{xspace}
\usepackage[printonlyused, nolist]{acronym}
\usepackage{lineno}

\newcommand{\Msun}{\ensuremath{M_{\odot}}\xspace}

\newcommand{\muBoundMin}{\ensuremath{1.3\times10^{-13}\,\mathrm{eV}}\xspace}
\newcommand{\muBoundMinNU}{\ensuremath{1.3\times10^{-13}}\xspace}
\newcommand{\muBoundMid}{\ensuremath{2\times10^{-13}\,\mathrm{eV}}\xspace}
\newcommand{\muBoundMax}{\ensuremath{2.7\times10^{-13}\,\mathrm{eV}}\xspace}
\newcommand{\muBoundMaxNU}{\ensuremath{2.7\times10^{-13}}\xspace}
\newcommand{\muBoundMinTight}{\ensuremath{2.2\times10^{-13}\,\mathrm{eV}}\xspace}
\newcommand{\muBoundMaxTight}{\ensuremath{2.7\times10^{-13}\,\mathrm{eV}}\xspace}
\newcommand{\muRef}{\ensuremath{10^{-12}\,\mathrm{eV}}\xspace}
\newcommand{\faMin}{\ensuremath{10^{14}~\mathrm{GeV}}\xspace}
\newcommand{\BFleft}{\ensuremath{0.5^{+0.1}_{-0.2}}\xspace}
\newcommand{\BFgap}{\ensuremath{5^{+5}_{-5}\times10^{-3}}\xspace}
\newcommand{\BFpeak}{\ensuremath{11.5^{+2.2}_{-1.3}}\xspace}
\newcommand{\BFfull}{\ensuremath{7.3^{+1.4}_{-1.1}\xspace}}

\begin{document}
\title{Constraints on Ultralight Scalar Bosons within Black Hole Spin Measurements from LIGO-Virgo's GWTC-2}
\author{Ken~K.~Y.~Ng}
\email{kenkyng@mit.edu}
\affiliation{LIGO Lab, Department of Physics, and Kavli Institute for Astrophysics and Space Research, Massachusetts Institute of Technology, 77 Massachusetts Avenue, Cambridge MA 02139, USA}
\author{Salvatore Vitale}
\affiliation{LIGO Lab, Department of Physics, and Kavli Institute for Astrophysics and Space Research, Massachusetts Institute of Technology, 77 Massachusetts Avenue, Cambridge MA 02139, USA}
\author{Otto~A.~Hannuksela}
\affiliation{Nikhef -- National Institute for Subatomic Physics, Science Park, 1098 XG Amsterdam, The Netherlands}
\affiliation{Department of Physics, Utrecht University, Princetonplein 1, 3584 CC Utrecht, The Netherlands}
\author{Tjonnie~G.~F.~Li}
\affiliation{Department of Physics, The Chinese University of Hong Kong, Shatin, NT, Hong Kong}
\affiliation{Institute for Theoretical Physics, KU Leuven, Celestijnenlaan 200D, B-3001 Leuven, Belgium}
\affiliation{Department of Electrical Engineering (ESAT), KU Leuven, Kasteelpark Arenberg 10, B-3001 Leuven, Belgium}
\date{\today}
\begin{abstract}
Clouds of ultralight bosons -- such as axions -- can form around a rapidly spinning black hole, if the black hole radius is comparable to the bosons' wavelength.
The cloud rapidly extracts angular momentum from the black hole, and reduces it to a characteristic value that depends on the boson's mass as well as on the black hole mass and spin.
Therefore, a measurement of a black hole mass and spin can be used to reveal or exclude the existence of such bosons.
Using the black holes released by LIGO and Virgo in their GWTC-2, we perform a simultaneous measurement of the black hole spin distribution at formation and the mass of the scalar boson.
We find that the data strongly disfavor the existence of scalar bosons in the mass range between $\muBoundMin$ and $\muBoundMax$.
Our mass constraint is valid for bosons with negligible self-interaction, that is with a decay constant $f_a\gtrsim\faMin$.
The statistical evidence is mostly driven by the two {binary black holes} systems GW190412 and GW190517, which host rapidly spinning black holes.
The region where bosons are excluded narrows down if these two systems merged shortly ($\sim 10^5$ yrs) after the black holes formed.
\end{abstract}
\maketitle

\section{Introduction}

Ultralight bosons are hypothetical particles with masses smaller than $\sim 10^{-11}$~eV.
Their existence, if verified, would help in solving open problems in particle physics and cosmology~\cite{PhysRevLett.38.1440,PhysRevLett.40.279,Preskill:1982cy,Abbott:1982af,Dine:1982ah,peccei2008strong,Bertone:2004pz,Arvanitaki:2009fg,Arvanitaki:2010sy,Marsh:2015xka,Hui:2016ltb}.
In fact, the name ultralight boson is commonly used to refer to multiple possible candidates, including fuzzy dark matter~\cite{hu2000fuzzy,Hui:2016ltb,Schive:2014dra}, dilatons~\cite{1996PhLB..379..105D,PhysRevD.82.084033,Arvanitaki:2014faa} and axions~\cite{PhysRevLett.38.1440,PhysRevD.16.1791,PhysRevLett.40.223,PhysRevLett.40.279,peccei2008strong,Cardoso:2018tly}.
Searches for ultralight bosons using tabletop experiments as well as astrophysical observations have been ongoing for years, covering decades of boson mass~\cite{PhysRevLett.105.171801,PhysRevLett.104.041301,PhysRevLett.105.051801,PhysRevLett.107.261302,Pugnat:2013dha,PhysRevD.95.083512,PhysRevD.96.061102,PhysRevLett.118.261301,PhysRevLett.118.061302,PhysRevLett.121.091802,2018QS&T....3a4008G,PhysRevLett.122.121802,2019PhRvL.123b1102D,Stadnik:2014tta,Stadnik:2015xbn,Dev:2016hxv,Abel:2017rtm,Arvanitaki:2016qwi,Brito:2017wnc,Brito:2017zvb,Isi:2018pzk,Hannuksela:2018izj,Tsukada:2018mbp,Morisaki:2018htj,Grote:2019uvn,berti2019ultralight,Palomba:2019vxe,Fernandez:2019qbj,baumann2019probing,Baumann:2019ztm,Kavic:2019cgk,XENON1T,Davoudiasl:2019nlo,Cunha:2019ikd,Sun:2019mqb,Martynov:2019azm,Siemonsen:2019ebd,CalderonBustillo:2020srq,Zhu:2020tht,Dergachev:2020fli,Annulli:2020ilw,Annulli:2020lyc,Tsukada:2020lgt,Michimura:2020vxn,Tsuchida:2020sms,Miller:2020vsl,Ng:2020jqd,Zhu:2020tht,CalderonBustillo:2020srq}.
To date, multiple constraints have been reported from nondetections~\cite{Tanabashi:2018oca}, together with a potential axion candidate from the XENON1T experiment~\cite{XENON1T}.
Gravitational-wave (GW) measurements of black holes in binaries (BBHs) provide a unique opportunity to detect or rule out the existence of these ultralight bosons in a mass range which is commensurate to the black holes masses and not accessible by lab-based experiment.
If such bosons exist and if their Compton wavelengths are comparable to the radius of a rapidly spinning black hole, boson superradiance may take place and generate a hydrogen-atom-like cloud around the spinning black hole~\cite{1971JETPL..14..180Z,Press:1972zz,bardeen1972,Dolan:2007mj,Arvanitaki:2009fg,Arvanitaki:2010sy,Brito:2014wla,Brito:2015oca,East:2018glu,Brito:2020lup}.
The cloud efficiently spins down the black hole to a characteristic critical spin, which depends on the boson mass, through a process called superradiant instability~\cite{Dolan:2007mj,Arvanitaki:2009fg,Arvanitaki:2010sy,Brito:2014wla,Brito:2015oca,East:2018glu,Brito:2020lup}.
Accessing tens or hundreds of BBHs thus allows for statistical tests on the existence of ultralight bosons, in a boson mass range that depends on the mass range of the population of black holes being probed~\cite{Arvanitaki:2009fg,Arvanitaki:2010sy,Arvanitaki:2014wva,Brito:2015oca,Arvanitaki:2016qwi,Brito:2017zvb,Brito:2017wnc,Baryakhtar:2017ngi,Stott:2017hvl,Stott:2018opm,Isi:2018pzk,antonio2018cw,Ghosh:2018gaw,Tsukada:2018mbp,Hannuksela:2018izj,berti2019ultralight,Palomba:2019vxe,Fernandez:2019qbj,baumann2019probing,Baumann:2019ztm,Kavic:2019cgk,Ng:2020jqd,Zhu:2020tht,Tsukada:2020lgt}.
For example, the stellar mass ($\sim 5$ to $\sim100$~\Msun) black holes that have been discovered by the ground-based GW detectors LIGO~\cite{aligo} and Virgo~\cite{virgo} can be used to probe boson masses in the range $3\times10^{-14}\,\mathrm{eV}\lesssim\mu_s\lesssim 10^{-11}\,\mathrm{eV}$~\cite{Arvanitaki:2014wva,Arvanitaki:2016qwi,Baryakhtar:2017ngi,Brito:2017zvb}.
Supermassive black holes, such as M87, can be used to probe much lighter bosons, with $\mu_s\sim10^{-21}$~eV~\cite{Davoudiasl:2019nlo}.
Roughly speaking, if a dearth of highly spinning black holes is observed for some range of black hole masses, that could be suggestive of the existence of ultralight bosons which have spun down the black holes.
Conversely, the discovery of highly spinning black holes could rule out the existence of a boson in an appropriate mass range.
This simple idea is made more complicated by a few factors.
First, one must take into account that some black holes may be slowly spinning \emph{when they form}.
The small spin measurements inferred from the BBH mergers observed by LIGO/Virgo could be due to either the superradiant growth of the boson cloud or an astrophysical distribution favoring small spins at the formation.
Reference~\cite{Ng:2019jsx} presented a Bayesian analysis where both the distribution of black hole spins at formation and the mass of the boson are considered, thus properly accounting for their correlation.
In this Letter we apply the methodology described in Ref.~\cite{Ng:2019jsx} by including the 45 binary black holes reported by the LIGO-Virgo-Kagra (LVK) Collaboration at high significance~\footnote{We follow Ref.~\citep{GWTC2rate} and only select the candidates with the false-alarm-rate (FAR) $<1\,\rm{yr}^{-1}$.} in Ref~\cite{GWTC2}.
We find the probability of a scalar boson with masses lying in the range $\muBoundMin\leq\mu_s\leq\muBoundMax$ is smaller than 0.01\%. The evidence against the existence of bosons with this mass arises mainly from two highly spinning black holes found in the new dataset, namely GW190412~\cite{GW190412} and GW190517.

\section{Constraints from GWTC-2}

We apply the Bayesian hierarchical method presented in Ref.~\cite{Ng:2019jsx} to all of the black holes reported by the LVK Collaboration in GWTC-1 and GWTC-2~\cite{GWTC1,GWTC2,gwosc,gwoscdata}~\footnote{We exclude the double neutron stars (NS) binaries GW170817 and GW190425, as well as the possible NSBH GW190426.
GW190719 and GW190909 are also excluded as their FARs are larger than $1\,\rm{yr}^{-1}$~\cite{GWTC2}.
}.
A detailed description of the method can be found in Ref.~\cite{Ng:2019jsx} and here we only summarize the main points. The main outcome of this analysis is a joint posterior for the distribution of the boson mass and the distribution of the black hole spins \emph{at formation}.
It is important to take into account the distribution of spins at formation, since the superradiant extraction of the spin angular momentum depends on the black hole properties and the boson mass.
Therefore, the fraction of black holes in the population that can undergo superradiance depends on the spin distribution at formation.
Following Ref.~\cite{Ng:2019jsx}, we use a beta distribution $p(\chi_F | \alpha,\beta)\propto \chi_F^{\alpha}(1-\chi_F)^{\beta}$ as our phenomenological model for the distribution of the formation spin $\chi_F$.
This distribution can capture some common configurations, such as a uniform $(\alpha=\beta=0)$ or a volumetric $(\alpha=2,\beta=0)$ distribution for the spin magnitude~\cite{Vitale:2017cfs}.
When $\alpha>\beta$ the beta distribution has more support for $\chi_F>0.5$, implying that more black holes are born with large spins and can be superradiantly spun down, making the inference of $\mu_s$ easier. The opposite is true for $\alpha<\beta$. 
In our analysis, we treat $\alpha$ and $\beta$ as additional free parameters, that are sampled together with $\mu_s$. Later, we marginalize the three-dimensional posterior $p(\mu_s,\alpha,\beta|\;\pmb{\rm d})$ over $(\alpha,\beta)$ to obtain the posterior for $\mu_s$.
These two parameters share the same prior, uniform in log in the range $[0.1,10]$.
We mention that the joint posterior of $(\alpha,\beta)$ is also interesting, as it carries information about the spin distribution at formation (see Fig.~4 of Ref.~\cite{Ng:2019jsx}). However, given the limited number of sources in GWTC-2, the inferred spin distribution at formation is not different from the spin distribution at merger as reported by Ref.~\cite{GWTC2rate}, and we thus do not report it here explicitly.

Another important factor to assess if black holes will be spun down by boson clouds is the time interval between the formation of the black hole and the merger: even if bosons of the appropriate mass exist, the black holes might not have the time to undergo superradiance when they merge too quickly after their birth.
As in Ref.~\cite{Ng:2019jsx}, we assume an inspiral timescale of $10$~Myr from the time the binary black hole system is formed to the time the black holes merge. This timescale is a conservative lower bound in light of population-synthesis studies~\cite{2000ApJ...528L..17P,MillerLauburg:2008,2009MNRAS.395.2127O,2011MNRAS.416..133D,2012PhRvD..85l3005K,2013ApJ...777..103T,2014MNRAS.441.3703Z,2015PhRvL.115e1101R,2016PhRvL.116b9901R,2015ApJ...800....9M,dominik2013double,Bavera:2019fkg}.
Since the inspiral timescale is usually much larger than the time it takes for a giant star to form a black hole, we assume that the two black holes in the binary are born simultaneously, and thus the inspiral timescale is a good probe for the lifetime of the individual black holes in the binary.

For the priors on black hole masses, we fix the BBH mass distribution to a power law for the mass of primary (heavier) black hole $M_1^{-2.35}$ within $[5,75]\,\Msun$ and a uniform distribution for the mass ratio $0.125\leq M_2/M_1\leq1$, consistent with the latest inferred population properties reported by the LVK collaboration~\cite{GWTC2rate}.

\begin{figure}[h]
    \centering
    \includegraphics[width=0.9\columnwidth]{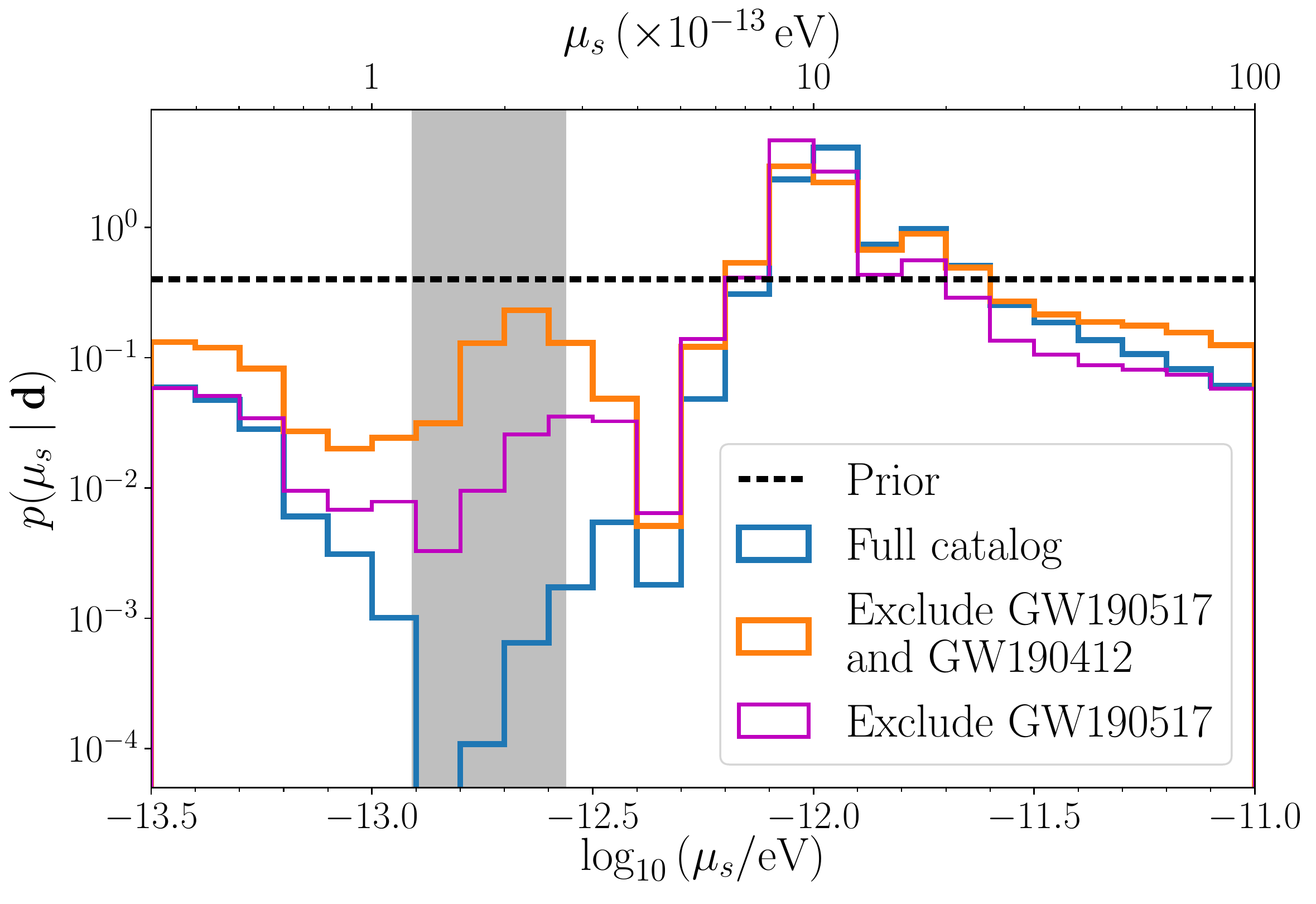}
    \caption{Marginalized posteriors (solid lines) of the scalar boson mass $\mu_s$ inferred from the dataset $\pmb{\mathrm{d}}$ consists of the full BBH catalog (blue), the dataset excluding GW190517 only (purple), as well as both GW190412 and GW190517 (orange).
    When the rapidly spinning BBHs GW190412 and GW190517 are included, there is only $0.01\%$ posterior support between $\muBoundMin\leq\mu_s\leq\muBoundMax$ (grey region).
    The prior (black dashed line) of $\mu_s$ is log uniform between $3\times10^{-14}$~eV and $10^{-11}$~eV.
    }
    \label{fig:bosonmass}
\end{figure}
Figure~\ref{fig:bosonmass} shows the marginalized posterior distribution for the boson mass inferred from the full BBH catalog (blue solid line).
A region with vanishing posterior support is clearly visible between $\muBoundMin$ and $\muBoundMax$: less than 0.01\% of the overall posterior is contained in this region, suggesting that the GWTC data strongly disfavour the existence of a boson within this narrow mass range.
Since large black hole spins at merger are at odds with the formation of boson clouds, this exclusion region must be caused by highly spinning black holes in the catalog.
Indeed, there are two primary black holes in GWTC-2 which are consistent with having large spin values: GW190412 and GW190517.
To check if the drop of posterior support evident in Fig.~\ref{fig:bosonmass} is caused by these two systems, we repeat the analysis by excluding GW190517 only (purple), as well as both GW190517 and GW190412 (orange).
Indeed the posterior of the boson mass using all sources but GW190412 and GW190517 does not show the same feature, and is instead much closer to the Bayesian prior we used (black dashed line).

\begin{figure}[h]
    \centering
    \includegraphics[width=0.9\columnwidth]{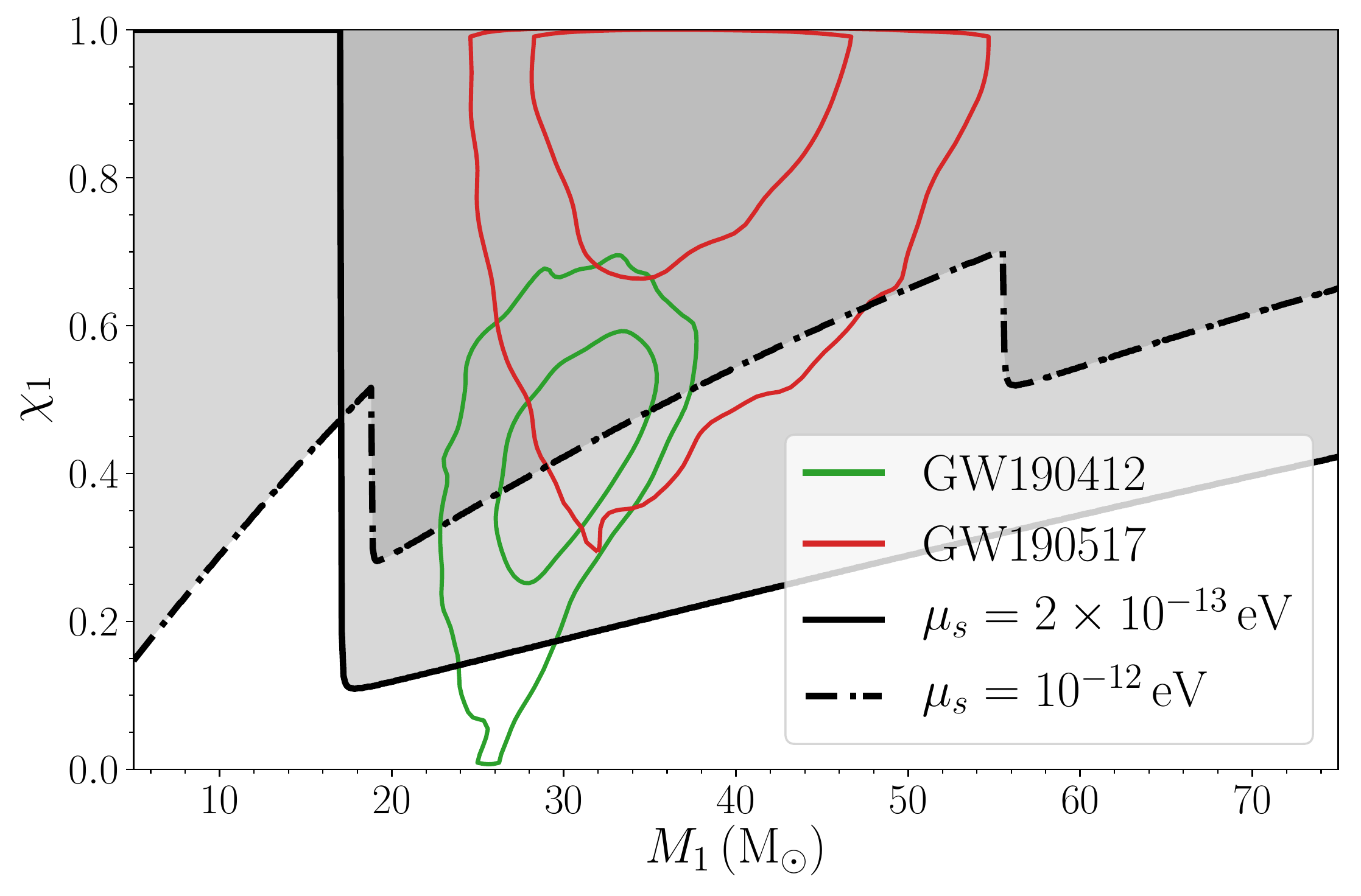}
    \caption{Exclusion regions (grey shaded region) enclosed by the critical spin curves of $\mu_s=\muBoundMid$ (black solid line) and $\mu_s=\muRef$ (black dash-dotted line) in the black hole mass-spin $(M_1,\chi_1)$ plane.
    The joint posteriors of the primary black holes of GW190412 (green contours) and GW190517 (red contours) are shown at 68\% and 95\% credible contours using the GWTC-2 default prior~\cite{GWTC2}.
    }
    \label{fig:reggeplane}
\end{figure}
To better understand how the spin measurements of GW190412 and GW190517 help excluding the existence of bosons, we overlay the joint mass-spin posteriors of the primary black hole in these two systems on the exclusion region generated by a boson with $\mu_s=\muBoundMid$, Fig.~\ref{fig:reggeplane}. The black solid line indicates the maximum postsuperradiance spin that a black hole could have as a function of its mass if a boson of mass $\mu_s=\muBoundMid$ existed: spins above the line (i.e in the grey region) are forbidden. 

We see that both of the primary black hole mass-spin posteriors have large overlaps with the exclusion region. In particular, the 95\% credible contour of GW190517 is entirely contained in the exclusion region for $\mu_s=\muBoundMid$, meaning that the primary black hole of GW190517 is inconsistent with having been spun down by the boson of this mass, hence heavily weighing down the existence of boson with mass $\mu_s=\muBoundMid$.
Different boson masses result in different exclusion regions: for example in Fig.~\ref{fig:reggeplane} we report the exclusion regions for a boson with mass $\mu_s=\muRef$ with a black dash-dotted line.
In this case, there is a non-negligible fraction of each posterior ($\sim 50\%$ and $\sim 5\%$ for GW190412 and GW190517, respectively) lying outside the exclusion region of $\mu_s=\muRef$.
This is why Fig.~\ref{fig:bosonmass} shows that the posterior for the boson mass is not vanishing for this value of the boson mass.

One's belief on a particular model (in this case, the existence of a boson with mass in some range) can be quantified using Bayesian model selections. We perform the analysis described in Ref.~\cite{Ng:2019jsx}
and calculate the Bayes factor between the ``boson model'' and the ``astrophysical model'' (that is, a model where there is no boson that sets off the process of superradiance. In this model the black hole spins are entirely determined by astrophysical processes).
Using a log uniform prior on $\mu_s$ between $\muBoundMax$ and $10^{-11}$~eV (that is, on the right of the grey band  visible in Fig.~\ref{fig:bosonmass}), we find a Bayes factor of $\BFpeak$ in favor of the boson model. While positive, this is much smaller than the threshold usually invoked for a strong statistical significance, i.e., $\geq 100$~\cite{ly2016harold}.
Hence, the data are inconclusive about the existence of bosons with mass $\mu_s>\muBoundMax$.
On the other hand, the Bayes factor for boson masses within the grey band in Fig.~\ref{fig:bosonmass}), i.e. in the range $[\muBoundMinNU,\muBoundMaxNU]$~eV, is $\BFgap$, smaller than the threshold $0.01$ and thus disfavoring the existence of a boson within this mass range.
In Tab.~\ref{tab:BFs} we also report the Bayes factor for boson with masses in the whole prior range, and with masses in the range $[3.16\times 10^{-14}, \muBoundMinNU]$, finding that in both cases the data are not informative.

\begin{table}
\begin{ruledtabular}
\caption{Bayes factors between the boson model and the astrophysical model for different ranges of $\mu_s$.
Larger values favor the boson model.}
\label{tab:BFs}
\begin{tabular}{cc}
 Range of $\mu_s$ (eV) & Bayes factor~\footnote{\label{col:BF} For each value, we report the medians and the 68\% credible intervals estimated from 50 nested-sampling chains.}\\
 \hline
 $[3.16\times 10^{-14}, \muBoundMinNU]$ & \BFleft\\
 $[\muBoundMinNU, \muBoundMaxNU]$ & \BFgap\\
 $[\muBoundMaxNU,10^{-11}]$ & \BFpeak\\
 $[3.16\times 10^{-14}, 10^{-11}]$ & \BFfull
\end{tabular}
\end{ruledtabular}
\end{table}

The appearance of a posterior excess around $\muRef$ in Fig.~\ref{fig:bosonmass} can be explained as follows.
If a boson of this mass existed, one would thus expect clustering of black hole spins along the critical spin curve (e.g. the solid and dot-dashed lines in Fig.~\ref{fig:reggeplane}), as well as a dearth of spins above the line.
The exact distribution depends on the boson mass which draws the critical spin curve; and the spin distribution at formation which determines the amount of black holes that can undergo superradiant spin down.
Therefore, as mentioned above, the posteriors on the spin distribution at formation and the boson mass are correlated (cf Ref.~\cite{Ng:2019jsx}).
The peak at $\muRef$ can thus be explained because, for that value of the boson mass, one would obtain black hole spins at merger which are similar (within a rather large uncertainty) to what is measured in the BBH dataset without invoking the existence of a boson.
With the current dataset, the algorithm cannot distinguish between a situation where black hole spins at formation are mostly small and bosonic clouds do not form, and the one where large amount of black hole have high spins at formation such that a boson with mass $\muRef$ exists and spins the black holes down. 

Owing to the lack of extensive numerical simulations on boson self-interaction, we do not allow for that possibility in our boson model.
Self-interaction would introduce nonlinear effect such as level mixing and ``bosenova''~\cite{Arvanitaki:2010sy,Yoshino:2012kn,Yoshino:2014wwa,Baryakhtar:2020gao}, and, if sufficiently large, it would stop the cloud growth before the saturation of superradiance (i.e. before the black hole spin has reached the critical spin).
As a result, the postsuperradiance spin might not decrease to the critical spin and be consistent with a large spin measurement.
The extent of the self-interaction is inversely proportional to the decay constant of the boson, $f_a$, and nonlinear effects become significant when the boson field reaches a maximum amplitude which depends on the black hole mass, the boson mass and the decay constant~\cite{Yoshino:2012kn,Yoshino:2014wwa,Baryakhtar:2020gao}.
Thus, we may use the mass measurement of the black holes that yield the $\mu_s$ constraint to estimate the value of $f_a$ above which the self-interaction is negligible~\cite{Arvanitaki:2010sy,Yoshino:2014wwa,Baryakhtar:2020gao}. Taking, for example, GW190517 (GW190412 has a similar primary mass and would thus yield a similar bound) -- i.e. $M_1\sim 35\,\Msun$ -- and using the nonlinear condition in Eq.~(7) of Ref.~\cite{Yoshino:2014wwa} with a typical energy for the boson cloud ($\sim10\%$ of the host black hole mass), we obtain that our analysis is certainly valid for $f_a\gtrsim\faMin$, which roughly includes the Grand-Unification-Theories energy scale for the constrained boson mass $\mu_s\approx \muBoundMid$~\cite{Arvanitaki:2010sy}.

\section{Discussion}

In this Letter, we have shown that the BBHs observed by LIGO and Virgo strongly disfavor the existence of scalar ultralight bosons with masses in the range $\muBoundMin\leq\mu_s\leq\muBoundMax$.
The statistical evidence is entirely contributed by the two highly spinning primaries in the systems GW190412 and GW190517.

Our method consistently accounts for the uncertainty of the black hole spin distribution at formation, which is marginalized over to obtain a posterior on the boson mass, Fig.~\ref{fig:bosonmass}.

However, caution is required in interpreting the results, since there are astrophysical scenarios that may explain the observed data without ruling out the existence of a boson  in that mass range.
The first caveat is related to the timescale between the formation of the black hole(s) and the merger of the binary, which has to be larger than superradiant timescale for a boson cloud to form and spin down the black hole in the first place.
As mentioned above, we assumed that the black holes lifetime is the same as the inspiral timescale, and took that to be 10~Myr, as suggested by simulation studies~\cite{2000ApJ...528L..17P,MillerLauburg:2008,2009MNRAS.395.2127O,2011MNRAS.416..133D,2012PhRvD..85l3005K,2013ApJ...777..103T,2014MNRAS.441.3703Z,2015PhRvL.115e1101R,2016PhRvL.116b9901R,2015ApJ...800....9M,dominik2013double,Bavera:2019fkg}.
This choice may not be valid if either of the GW190412 or GW190517 binaries was formed with an extremely high eccentricity $1-e\lesssim0.01$ shortly after the birth of the component black holes, such that their inspiral timescales are reduced by few orders of magnitude~\cite{Peters:1964zz,Wen2005}.
In this scenario, there would not be time for black holes to lose their spin to superradiance, and they may retain large spins even if a boson exists, reducing the significance of our constraints.
Production of extremely eccentric BBHs is possible in dense stellar clusters or active galactic nuclei (AGN), but these BBHs with extreme eccentricity are expected to have very low merger rates~\cite{Rodriguez:2018pss,Grobner:2020drr,Samsing:2020tda,Martinez:2020lzt}.
The AGN environment may also enhance the production of hierarchical binaries, i.e., binaries made of previous merger remnants, that merge in a very short timescale $\sim10^5$~yr~\cite{Bartos:2016dgn,Yang:2019cbr}.
Assuming this shorter timescale as the black hole lifetime, we find that the exclusion range of boson masses narrows to $\muBoundMinTight\leq\mu_s\leq\muBoundMaxTight$.

The second caveat is related to the possible gas accretion onto the black holes, which we have ignored in this work.
The black hole spin gradually increases when the materials of the rotating accretion disk keep falling into the black hole.
The evolution of the black hole spin thus depends on the how significant the accretion can be. If the \textit{spin-up} rate due to accretion is much faster than the \textit{spin-down} rate due to superradiance, then the black holes may end up having a large spin, inside the exclusion region, even if bosons exist exists.
In the opposite case, superradiant spin-down dominates and the black holes should still ends its life with a spin around the critical spin curve.
For the stellar mass black holes relevant for ground-based GW detectors, even an accretion rate at the Eddington limit is expected to be much smaller than the typical superradiant rate~\cite{Brito:2015oca,Brito:2017zvb,Sun:2019mqb}.
Therefore, our results are still robust unless there is a thin-disk accretion whose rate is drastically and continuously super-Eddington throughout the black hole lifetime~\cite{Bardeen:1970zz,King:1999aq}.
This is unlikely to be the case for binary black holes even in gas rich astrophysical environments, but not strictly impossible~\cite{Yang:2019cbr,Yi:2019rwo,vanSon:2020zbk}.

The gravitational potential of the companion in a BBH may alter the superradiant growth due to tidal interaction.
However, the tidal disruption may excite the in-falling modes with opposite angular momentum and is likely to enhance the spin-down of the host black hole~\cite{baumann2019probing,Baumann:2019ztm,berti2019ultralight}, and may further broaden the exclusion regions~\cite{Ficarra:2018rfu}.
We also note that the mass loss due to superradiance is ignored, which contributes to a few percent overestimation of the boson mass constraints~\cite{Ficarra:2018rfu,Isi:2018pzk,Ng:2019jsx}.

The constraints presented in this Letter will improve in the future, if the spins of heavier black holes are found to be above their critical spin curve.
Second-generation black hole mergers, whose primary black holes have a spin at formation $\chi\sim0.7$ and large masses - $M\gtrsim 50\,\Msun$~\cite{Gerosa:2017kvu,Fishbach:2017dwv,Kimball:2020opk} - might be the ideal candidates to test for the existence of lighter boson, $\mu_s\lesssim 10^{-13}$~eV, with ground-based GW detectors.
On the other hand, if a boson existed with mass $\mu_s\approx\muRef$, for which we have found weak evidence, its existence could be shown with a few more hundred more black-hole spin measurements, needed to verify the clustering of black hole spins along the corresponding critical spin curve (dot-dashed line in Fig~\ref{fig:reggeplane}~\cite{Ng:2019jsx}).
We end by remarking that constraints on ultralight bosons with GWs can also be obtained by targeting the nearly monochromatic GWs emitted by the cloud of bosons~\cite{Palomba:2019vxe,Tsukada:2018mbp,Sun:2019mqb,Zhu:2020tht,Dergachev:2020fli,Ng:2020jqd}. The two approaches target black holes at different stages of their life. In particular, the method based on continuous waves requires the cloud to be present at the time of the measurement, while the approach described in this Letter focuses on the black holes after they have been spun down.
These two approaches also use entirely different statistical methods, therefore yielding complementary constraints. 

\section{Acknowledgements}
We thank Juan Calderon Bustillo, Will Farr, Hartmut Grote, Max Isi, and Lilli Sun for valuable discussions and suggestions.
K. K. Y. N. and S. V., members of the LIGO Laboratory, acknowledge the support of the National Science Foundation through the NSF Grant No. PHY-1836814. LIGO was constructed by the California Institute of Technology and Massachusetts Institute of Technology with funding from the National Science Foundation and operates under Cooperative Agreement No. PHY-1764464. O. A. H. is supported by the research program of the Netherlands Organization for Scientific Research (NWO). T. G. F. L. was partially supported by grants from the Research Grants Council of Hong Kong (Projects No. CUHK14306218, No. CUHK14310816, and No. CUHK24304317), Research Committee of the Chinese University of Hong Kong, and the Croucher Foundation in Hong Kong. The authors are grateful for computational resources provided by the LIGO Lab and supported by the National Science Foundation Grants No. PHY-0757058 and No. PHY-0823459. This research has made use of data, software and/or web tools obtained from the Gravitational Wave Open Science Center~\cite{gwoscdata}, a service of LIGO Laboratory, the LIGO Scientific Collaboration, and the Virgo
Collaboration.

\bibliography{reference}
\end{document}